\documentclass[preprint]{revtex4}
\usepackage{amsmath,amssymb,verbatim,graphicx}
\usepackage[paperwidth=210mm,paperheight=297mm,centering,hmargin=2cm,vmargin=2.5cm]{geometry}

\newcommand{\beginappendixA}{%
        \setcounter{table}{0}
        \renewcommand{\thetable}{A.\arabic{table}}%
        \setcounter{figure}{0}
        \renewcommand{\thefigure}{A.\arabic{figure}}%
        \setcounter{equation}{0}
        \renewcommand{\theequation}{A.\arabic{equation}}%
     }
\newcommand{\beginappendixB}{%
        \setcounter{table}{0}
        \renewcommand{\thetable}{B.\arabic{table}}%
        \setcounter{figure}{0}
        \renewcommand{\thefigure}{B.\arabic{figure}}%
        \setcounter{equation}{0}
        \renewcommand{\theequation}{B.\arabic{equation}}%
     }
     
\begin{document}

%%%%%%%%%%%%%%%%%%%%%%%%%%%%%%%%%%%%%%%%%%%%%%%%%%%%%%%%%%%%%%%%%%%%%%%%%%%%%%%%%%%%%
\title{Exploring classical correlations in noise to recover quantum information using local filtering}
\author{Daniel E. Jones,$^{1}$ Brian T. Kirby,$^{1}$ Gabriele Riccardi,$^{3}$ Cristian Antonelli,$^{3}$ and Michael Brodsky$^{1,2}$}% <-this % stops a space
\affiliation{\\$^{1}$ U.S. Army Research Laboratory, Adelphi, MD 20783, USA\\
$^{2}$ U.S. Military Academy, West Point, NY 10996, USA\\
$^{3}$ Department of Physical and Chemical Sciences, University of L’Aquila, L’Aquila 67100, Italy\\
E-mail: daniel.e.jones161.civ@mail.mil and michael.brodsky4.civ@mail.mil
}
%%%%%%%%%%%%%%%%%%%%%%%%%%%%%%%%%%%%%%%%%%%%%%%%%%%%%%%%%%%%%%%%%%%%%%%%%%%%%%%%%%%%%

%%%%%%%%%%%%%%%%%%%%%%%%%%%%%%%%%%%%%%%%%%%%%%%%%%%%%%%%%%%%%%%%%%%%%%%%%%%%%%%%%%%%%
\begin{abstract} 
A general quantum channel consisting of a decohering and a filtering element carries one qubit of an entangled photon pair.  
As we apply a local filter to the other qubit, some mutual quantum information between the two qubits is restored depending on the properties of the noise mixed into the signal. 
We demonstrate a drastic difference between channels with bit-flip and phase-flip noise and further suggest a scheme for maximal recovery of the quantum information. 
\end{abstract}
%%%%%%%%%%%%%%%%%%%%%%%%%%%%%%%%%%%%%%%%%%%%%%%%%%%%%%%%%%%%%%%%%%%%%%%%%%%%%%%%%%%%%

\maketitle

% npjQI brief comms DO NOT HAVE headings, but they DO HAVE a methods section
%%%%%%%%%%%%%%%%%%%%%%%%%%%%%%%%%%%%%%%%%%%%%%%%%%%%%%%%%%%%%%%%%%%%%%%%%%%%%%%%%%%%%
\section{Introduction}

Rapid advances in photonic quantum information physics and engineering are fueled by enticing promises of new computational and communication paradigms realizable in future quantum networks.
Recent roadmaps for a quantum internet necessarily encompass fast and reliable transfer of quantum information between distant locations \cite{wehner2018quantum, diamanti2016practical, kimble2008quantum, ecker2019overcoming}.
In the most general case, an entanglement distillation protocol \cite{bennett1996purification,deutsch1996quantum,pan2003experimental,abdelkhalek2016efficient} is believed to be able to ensure the error-free transmission of quantum states. 
Some of these distillation protocols can be viewed as the filtering of the transmitted quantum signal from some kind of decohering influence of various types of noise inherent to a communication system. 
% Separately, high-dimensional entanglement has also recently been used to distribute entanglement in the presence of noise \cite{ecker2019overcoming}.

Filtering is a known and powerful tool in the quantum engineer’s tool box. 
Besides entanglement distillation \cite{kwiat2001experimental,park2010construction}, local filters have been used for a number of important functions such as remote two-photon state preparation \cite{peters2005remote}, purification of individual qubits \cite{ricci2004experimental}, multi-dimensional state characterization \cite{howland2016compressively,friis2019entanglement}, increasing resilience to amplitude damping \cite{kim2012protecting,lee2014experimental}, correcting mode misalignment  \cite{asano2015distillation}, and estimating channel capacity \cite{cuevas2017experimental}.
Furthermore, spatial and modal filtering schemes were proposed to separate signal and noise \cite{huang2011distilling,layden2018spatial}.
More complex functions include realizing noiseless amplification \cite{chrzanowski2014measurement}, decoupling an eavesdropper from a lossy channel \cite{jacobsen2018complete}, and potentially removing bit-flipped errors using the entanglement filter \cite{okamoto2009entanglement}.

In our previous work \cite{jones2018tuning,kirby2019effect}, we addressed a curious interplay between two local filters applied to a perfect Bell state. 
One filter was realized by inherent imperfections of a quantum channel carrying one photon of a pair, and the other filter was deliberately controlled by a quantum network operator. 
For that scenario, we showed how the operator can trade off transmission rates for entanglement quality for the ideal condition of noiseless initial photon pairs. 
For a sufficiently narrow frequency range, a general quantum channel can be represented by a combination of an arbitrarily oriented decohering element, which mixes noise into the initial Bell state, followed by a local filter \cite{damask2004polarization,shtaif2005polarization}. 
The relative orientation of the decoherence and filtering elements determines what kind of noise is mixed into the signal.
Whether the entanglement-rate trade-off mentioned above works for the various types of partially mixed pairs is the subject of the current paper.

In this paper, we consider the transmission of one photon of a Bell state over a general quantum channel which adds noise and introduces inherent filtering that partially removes one mode. 
We experimentally investigate two exemplary cases of noise -- one corresponding to bit-flip errors and the other corresponding to phase-flip errors.
The errors and the subsequent filtering in the channel diminish the quality of distributed entanglement. 
We then employ the Procrustean filtering method \cite{bennett1996concentrating, thew2001entanglement,thew2001mixed,kwiat2001experimental}, that is we use a local filter on the remaining photon of the pair. 
By tuning the filter magnitude and direction, we recover as much of the mutual quantum information between the two photons as possible. 
We prove that for uncorrelated noise due to bit-flip errors, the Procrustean filtering method is very ineffective and can only recover a small amount of mutual information. 
The method can, however, be marginally improved by a judicious choice of the filter magnitude. 
Next, we explore classically correlated noise arising when the channel introduces phase-flip errors. 
We show that correlated noise does not hinder the ability to completely recover mutual information by Procrustean filtering.
Numerical extraction from theoretical density matrices validates our data.
Our experiments suggest that a polarization controller properly pre-positioned in mid-channel can enable successful recovery of quantum information.  

%%%%%%%%%%%%%%%%%%%%%%%%%%%%%%%%%%%%%%%%%%%%%%%%%%%%%%%%%%%%%%%%%%%%%%%%%%%%%%%%%%%%%
\section{Quantum State Preparation}

\begin{figure*}[!t]
\centering
\includegraphics[width=5.5in]{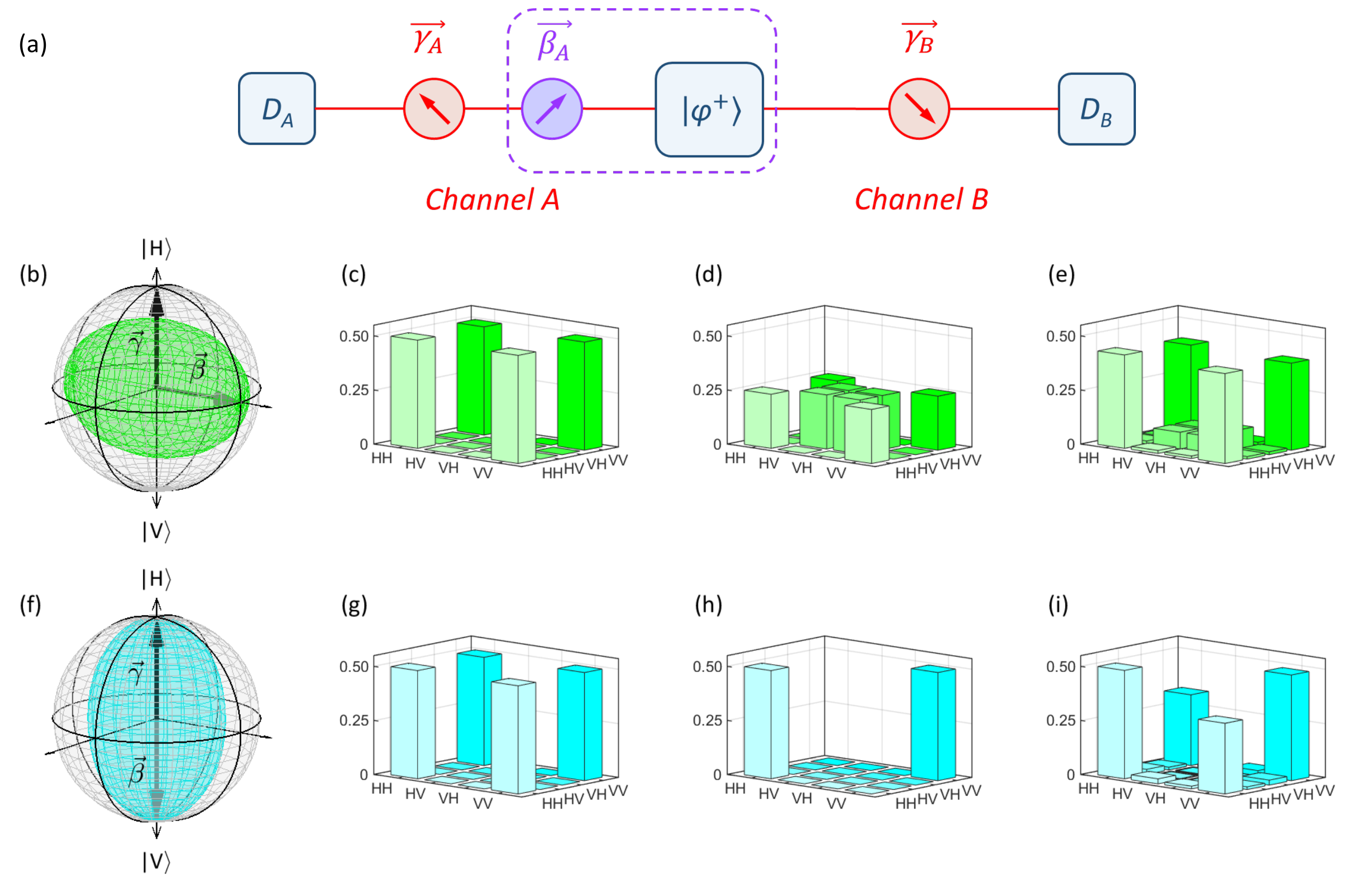}%width=\linewidth
\caption{(a) Experiment schematic: a source outputs a $|\phi^{+}\rangle$ state towards detectors $D_{A}$ and $D_{B}$. Channel A is a concatenation of a decoherence element $\vec{\beta}_{A}$ and a mode filter $\vec{\gamma}_{A}$, while channel B only includes a mode filter $\vec{\gamma}_{B}$. Vectors $\vec{\beta}_{A}$ and $\vec{\gamma}_{A}$ are perpendicular in (b, c, d, e) and collinear in (f, g, h, i). 
(b, f) Evolution maps of qubit A through $\vec{\beta}_{A}$. 
(c, g) Theoretical density matrices for the ``quantum signal" $|\phi^{+}\rangle$. 
(d, h) Theoretical density matrices for the ``quantum noise" introduced by $\vec{\beta}_{A}$: bit-flip noise $N_{\text{bf}}$ (green) and phase-flip noise $N_{\text{pf}}$ (cyan). 
(e, i) Experimentally obtained density matrices for $\vec{\gamma}_{A}=0$ and $\vec{\gamma}_{B}=0$. These are clearly incoherent mixtures of the signal and noise pictured to the left.
}
\label{fig:setup}
\end{figure*}

Our experiment is shown schematically in Fig. \ref{fig:setup}(a). 
Photons from a source outputting a $|\phi^{+}\rangle$ Bell State are separated and routed to quantum channels A and B. 
For channel A, representing interactions with the environment, we choose a general polarization quantum channel comprising a decohering birefringent element $\vec\beta_A$ of arbitrary orientation, followed by a mode filter inherent to the channel $\vec\gamma_{A}$. 
The direction of vectors $\vec\beta_A$ and $\vec\gamma_{A}$ on the Bloch sphere as well as the filter's magnitude are variable. 
Channel B represents the control by a quantum network operator and contains a mode filter $\vec\gamma_{B}$ that is also tunable in both magnitude and direction. 
The physical properties $\vec\gamma$ and $\vec\beta$ of our channels can be conveniently represented by rotational operators in Jones space \cite{damask2004polarization, gordon2000pmd}.
The operator for a filter $\vec\gamma$ is given by: $\textstyle P=\exp\left({\frac{\vec\gamma}{2}\cdot\vec\sigma}\right)$, where  $\vec\sigma$ is a vector whose elements are the three Pauli matrices.
%The operator for a filter $\vec\gamma$ is given by: $\textstyle P=\exp\left(-\frac{\gamma}{2}\right)\exp\left({\frac{\vec\gamma}{2}\cdot\vec\sigma}\right)$, where  $\vec\sigma$ is a vector whose elements are the three Pauli matrices.
Likewise, the rotational operator for a birefringent element $\vec\beta$ of length $L$ is: $\textstyle U=\exp\left({-j\frac{\vec\beta L}{2}\cdot\vec\sigma}\right)$.
The differential group delay between two pulses of the slow and fast polarization modes is equal to $\textstyle |\vec\tau|=|\frac{\partial\vec\beta}{\partial\omega}|$, which is a constant in our experiment. Setting and adjusting the decoherence element allows us to vary the noise admixed to our quantum state. 

To implement the system of Fig. 1(a), we generate signal and idler photon pairs via four-wave mixing \cite{fiorentino2002all} by pumping a dispersion shifted fiber (DSF) with a 50 MHz pulsed fiber laser centered at 1552.52 nm.
The DSF is arranged in a Sagnac loop with a polarization beam splitter (PBS) in order to entangle the signal and idler in polarization, and a WDM demux separates the photons spectrally into 100 GHz-spaced ITU outputs after the Sagnac loop, resulting in photons with a temporal duration of about 15 ps \cite{wang2009robust}. 
The pump pulse is also filtered by a 100 GHz telecom add/drop filter. 
ITU channel 28 (1554.94nm) is sent to channel A and ITU channel 34 (1550.12nm) to channel B.  
The source typically outputs an average number of pairs per pump pulse in the $\mu = 0.001 - 0.1$ range \cite{jones2017joint,jones2017situ}. 
Tunable polarization dependent loss emulators \cite{Liboiron2006polarization} serve as the filters $\vec\gamma_A$ and  $\vec\gamma_B$, while a fixed polarization mode dispersion emulator \cite{BRODSKY2008605} provides the decohering element $\vec\beta_A$. 
Polarization controllers set the orientation of the individual vectors $\vec\beta_A$, $\vec\gamma_A$, and $\vec\gamma_B$. 
The detectors operate in a gated mode with a detection efficiency of $\eta \sim 20\%$ and a dark count probability of $ \sim 4 \times 10^{-5}$ per gate. 
Automated FPGA-based controller software performs full polarization state tomography to determine the density matrix of the state measured at the two detector stations \cite{altepeter2005photonic, nucrypt}.

%%%%%%%%%%%%%%%%%%%%%%%%%%%%%%%%%%%%%%%%%%%%%%%%%%%%%%%%%%%%%%%%%%%%%%
%%%%%%%%%%%%%%%%%%%%%%%%%%%%%%%%%%%%%%%%%%%%%%%%%%%%%%%%%%%%%%%%%%%%%%

To illustrate how the birefringence $\vec\beta_A$ depolarizes the qubits traversing channel A, we turn to the Bloch sphere representation. 
In Fig. \ref{fig:setup}(b, f), we present the degree to which a sphere representing qubit A shrinks into a prolate spheroid aligned along the direction of $\vec\beta_A$. 
Note that the $|H\rangle$ direction for photon A is chosen along $\vec\gamma_A$.
In general, the orientation of $\vec\beta_A$ can be arbitrary, but here we study two characteristic cases. 
In the first case, the birefringence vector $\vec\beta_A$ points toward the equator as seen in Fig. \ref{fig:setup}(b). 
The resulting density matrix can be described as a bit-flipping channel with probabilistic action on qubit A given by $\rho\rightarrow\rho' =(1- \textstyle \frac{p}{2} ) \rho + \textstyle \frac{p}{2} \sigma_1\rho\sigma_1$. 
Figure \ref{fig:setup}(f) shows the second case, in which the birefringence vector $\vec\beta_A$ is oriented toward the pole. 
The polar birefringence acts as a dephasing or phase-flipping channel, in which probabilistic evolution of the qubit A density matrix is given by $\rho\rightarrow\rho' =(1-\textstyle \frac{p}{2}) \rho + \textstyle \frac{p}{2} \sigma_3\rho\sigma_3$. 
Here, $\sigma_{1}$ and $\sigma_{3}$ are Pauli matrices and the $p$ value determines the amount of noise mixed into the state.
Note that the $p$ value used in the numerical plots of Fig. \ref{fig:setup}(b, f) matches the experimentally observed value of $p=0.33$. 

In addition to the probabilistic qubit rotations above, there are other ways of implementing the maps of Fig. \ref{fig:setup}(b, f). 
For instance, qubits containing a range of frequency components will be similarly dephased by a frequency dependent birefringence $\vec\beta_A$ \cite{Antonelli2011Sudden, brodsky2011loss, shtaif2011nonlocal}, which is the method that we have implemented here. 
To describe the overall effect of the channel A decoherence $\vec\beta_A$ on the maximally-entangled initial two-qubit $|\phi^{+}\rangle$ Bell state, we use Choi-Jamio\l{}kowski isomorphism \cite{choi1975completely,jamiolkowski1972linear}. 

Consider first the equatorial orientation of $\vec\beta_A$, that is the case of bit-flip errors. 
The two-qubit density matrix $\rho_{\text{bf}}$ becomes:
\begin{equation}
\rho_{\text{bf}}=(1-p)|\phi^{+}\rangle\langle\phi^{+}|+pN_{\text{bf}},
\label{eq:bitflip}
\end{equation}
where $N_{\text{bf}}=\textstyle\frac{1}{2}(|\phi^{+}\rangle\langle\phi^{+}|+|\psi^{+}\rangle\langle\psi^{+}|)$. 
The sum of the two terms, each illustrated for clarity in Figs. \ref{fig:setup}(c) and (d), represents an incoherent mixture of the original quantum signal and the rank two bit-flip noise admixed to it by the decoherence element. 
Experimental characterization of the state emerging at the output of the dashed purple box in Fig. \ref{fig:setup}(a) produces the density matrix $\rho_{\text{bf}}$ presented in Fig. \ref{fig:setup}(e). 
The matrix is clearly well described by Eq. \ref{eq:bitflip}, hence verifying our model and attesting to the precision with which we are able to control our quantum states.

The polar orientation of $\vec\beta_A$, that is the case of phase-flip errors, results in a very different noise term.
Again, the two-qubit density matrix $\rho_{\text{pf}}$ can be represented as an incoherent mixture of the quantum signal and the quantum noise:
\begin{equation}
\rho_{\text{pf}}=(1-p)|\phi^{+}\rangle\langle\phi^{+}|+pN_{\text{pf}},
\label{eq:phaseflip}
\end{equation}
but here $N_{\text{pf}}=\textstyle\frac{1}{2}(|\phi^{+}\rangle\langle\phi^{+}|+|\phi^{-}\rangle\langle\phi^{-}|)$. 
The two terms are shown in Figs. \ref{fig:setup}(g) and (h).
The actual experimentally measured density matrix closely follows Eq. \ref{eq:phaseflip} as can be seen in Fig. \ref{fig:setup}(i). 

Both noise terms $N_{\text{bf}}$ and $N_{\text{pf}}$ are rank-2 Bell diagonal states. 
Compared to the isotropic rank-4 noise theoretically considered previously for Werner states \cite{kirby2019effect}, the terms $N_{\text{bf}}$ and $N_{\text{pf}}$ represent more realistic scenarios. 
The properties of the two are drastically different, however. Note that $N_{\text{pf}}$ is a mixture of co-polarized classical photons pairs $|HH\rangle$ and $|VV\rangle$. Therefore, it exhibits classical correlations in the canonical basis, the significance of which we show below.

%%%%%%%%%%%%%%%%%%%%%%%%%%%%%%%%%%%%%%%%%%%%%%%%%%%%%%%%%%%%%%%%%%%%%%%%%%%%%%%%%%%%%
\section{Results}

We now investigate further entanglement degradation of the states of Eqs. \ref{eq:bitflip}-\ref{eq:phaseflip} by the inherent filtering effect of channel A, $\vec\gamma_{A}$, as well as the nonlocal entanglement restoration achieved by applying an additional filter $\vec\gamma_{B}$ in channel B. 
Here, we quantify the amount of entanglement by the mutual quantum information \cite{adami1997neumann} shared between qubits A and B. 
The mutual information can be readily computed from the experimentally obtained density matrices by using $S(A:B) = S(A) + S(B) - S(AB)$, where $S(AB)$ is the von Neumann entropy of the entire two-qubit state, and $S(A)$, $S(B)$ are the marginal entropies. 

\begin{figure*}[t]
\centering
\includegraphics[width=6.5in]{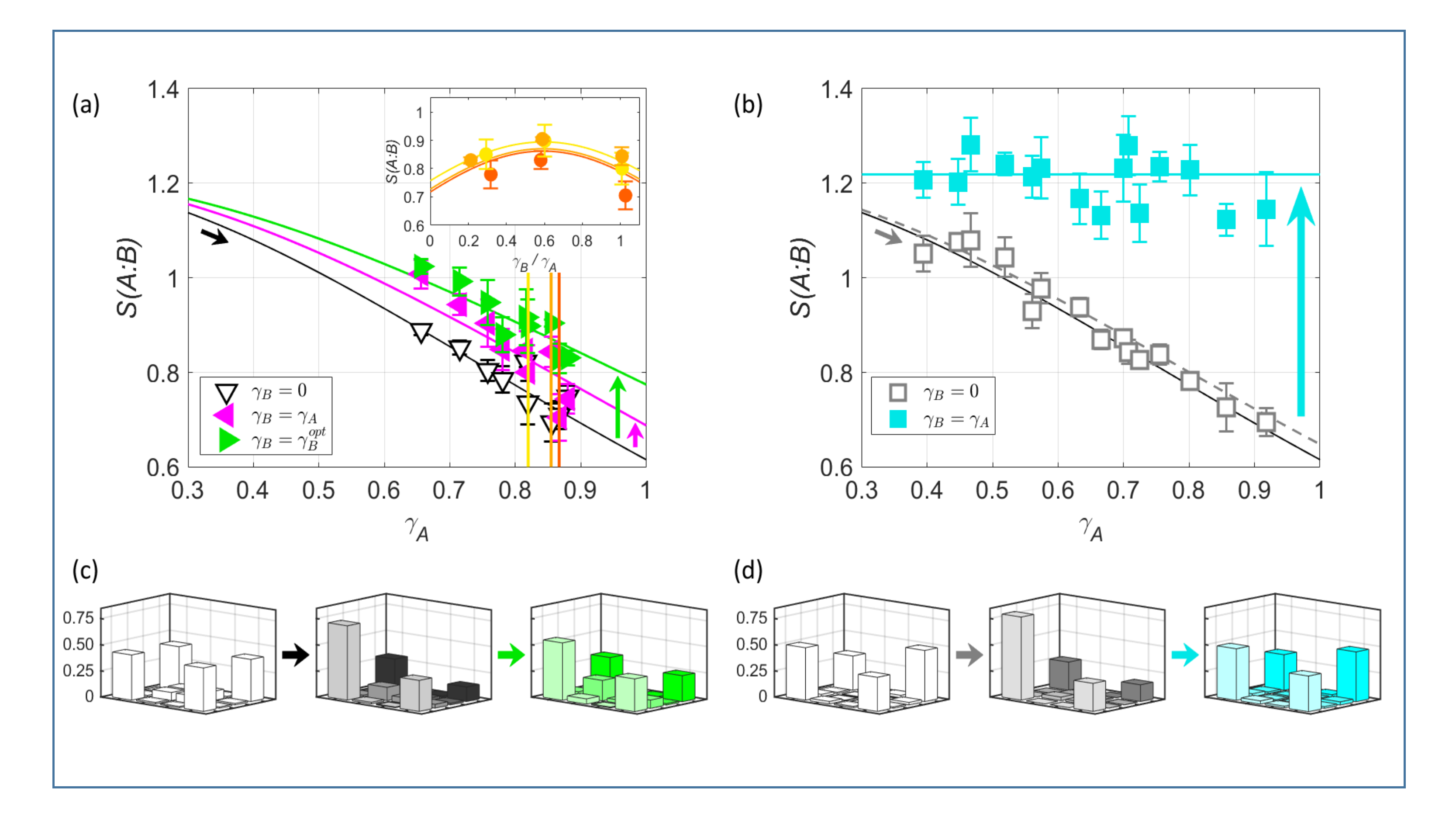}%width=\linewidth
\caption{
(a) Bit-flip noise. Experimental (symbols) and theoretical (lines) mutual information $S(A:B)$ as a function of $\gamma_{A}$. Symbols denote $\gamma_{B}$ values: $\gamma_{B}=0$ (empty $\triangledown$), $\gamma_{B}=\gamma_{A}$ (purple $\lhd$), and $\gamma_{B}=\gamma_{B}^{\text{opt}}$ (green $\rhd$). For the values of $\gamma_{A}$ denoted by the vertical lines, the mutual information $S(A:B)$ is shown vs the ratio of the two filter magnitudes $\frac{\gamma_{B}}{\gamma_A}$ (inset).
(b) Phase-flip noise. Experimental ($\Box$) and theoretical (lines) mutual information $S(A:B)$ as a function of $\gamma_A$. Symbols denote $\gamma_{B}$ values: $\gamma_{B}=0$ (empty $\Box$) and $\gamma_{B}=\gamma_{A}$ (cyan $\Box$). The black line from (a) is also shown for comparison.
(c) Bit-flip noise. Progression of experimental density matrices: initial mixed state after decoherence but no filtering (left), after filtering in channel A (middle), after compensation by filtering channel B (right). 
(d) Phase-flip noise. Progression of experimental matrices similar to (c).
The polarization bases of the density matrices in (c) and (d) are the same as those in Fig. 1 (c,d,e,g,h,i). 
}
\label{fig:data}
\end{figure*}

Figure \ref{fig:data}(a) shows a decrease in the measured $S(A:B)$ with empty black triangles when a filter in channel A, $\vec\gamma_{A}$, of increasing strength is applied to the bit-flipped state of Eq. \ref{eq:bitflip}. 
Some quantum information can be recovered by applying another filter $\vec\gamma_{B}$ to channel B. 
We first select its magnitude, $\gamma_{B}$, and then its orientation such that maximal information recovery is achieved. 
For the symmetric case of $\gamma_{B}=\gamma_{A}$, the best recovery achieved is shown in purple triangles; however, the maximum amount of recovered quantum information can be further increased by slightly reducing $\gamma_{B}$.
We determined that the optimal filter magnitude, $\gamma_{B}^{\text{opt}}$, is given by $\gamma_{B}^{\text{opt}}=\textrm{tanh}^{-1}[C \,\, \textrm{tanh}(\gamma_{A})]$, where $C$ is the concurrence of the bit-flipped state of Eq. \ref{eq:bitflip}.
The derivation of this optimal filter magnitude can be found in Appendix A.
Next, we adjusted the magnitude of the filter in channel B to the optimum value, $\gamma_{B} = \gamma_{B}^{\text{opt}}$.  
The results, shown with the green triangles in Fig. \ref{fig:data}(a), demonstrate the best possible information recovery for the bit-flipped state created by our setup. 
The scatter seen in the data is largely due to the inherently imperfect alignment of the various elements, whereas the uncertainty arising from photon counting is indicated by the error bars.
The lines show the corresponding normalized theory (see Appendix B for details).
The range of the axes in Fig. 2(a) was chosen to be the same as in Fig. 2(b) to facilitate comparison between the two cases.
Figure \ref{fig:data}(a) convincingly demonstrates that Procrustean filtering can only recover a small amount of the mutual information lost by quantum states corrupted by bit-flip errors.  

%%%%%%%%%%%%%%%%%%%%%%%%%%%%%%%%%%%%%%%%%%%%%%%%%%%%%%%%%%%%%%%%%%%%%%
%%%%%%%%%%%%%%%%%%%%%%%%%%%%%%%%%%%%%%%%%%%%%%%%%%%%%%%%%%%%%%%%%%%%%%

The inset of Fig. \ref{fig:data}(a) further illustrates the effect of the relative magnitudes of $\vec\gamma_{A}$ and $\vec\gamma_{B}$ on the amount of information which can be recovered. 
Here, we plot the measured mutual information $S(A:B)$ (symbols) and the corresponding theory (lines) as a function of the $\gamma_{B}/\gamma_{A}$ ratio. 
The data is taken at the three different values of $\gamma_{A}$ denoted by the vertical lines in the main panel, that is at $\gamma_{A} = .820, .857, .869$. 
The points and curves in the inset correspond to the points along the vertical lines of the same color in the main panel. 
Our theory gives a nearly equal optimal value for all three cases, $\gamma_{B}^{\text{opt}} \simeq 0.59 \,\gamma_{A}$. 
Both the data and numerical simulations shown on the inset clearly validate our theory. 

Figure \ref{fig:data}(b) presents data similar to that of Fig. \ref{fig:data}(a), but for the case of phase-flip noise. 
Similar to the black squares in Fig. \ref{fig:data}(a), the gray squares show the mutual information as a function of the inherent filter strength of channel A for the two-qubit state impaired by phase-flip noise.
The dashed gray line shows the theory for the phase-flip case, and the solid line taken from Fig. 2(a) for the bit-flip case is also plotted for comparison.
The mutual information between two qubits decreases somewhat slower for the phase-flipped case as $\gamma_{A}$ is increased. 
This is a surprising observation since the concurrence as a function of $\gamma_{A}$ is exactly the same in both cases. 
That is, the relative orientation of the filter $\vec\gamma_{A}$ with respect to the decoherence $\vec\beta_{A}$ changes the amount of information shared between the two qubits without changing the degree of entanglement of the pair.
We speculate that the slower decrease of the mutual information is due to existing correlations in the phase-flip noise itself. 

Similar to our procedure for recovering the information lost in the bit-flip case, we then applied an additional filter $\vec{\gamma}_B$ to recover the information lost due to $\vec{\gamma}_A$.
Here, the symmetric case $\gamma_{B}=\gamma_{A}$ allows for the recovery of all information lost due to $\vec{\gamma}_A$, as in the case of a perfectly entangled state \cite{jones2018tuning}. 
The cyan squares show the resulting mutual information measured after the application of the channel B filter with magnitude $\gamma_{B}=\gamma_{A}$. 
The cyan line corresponds to the mutual information when $\gamma_{A}=0$, confirming that $\vec{\gamma}_B$ compensates for the effect of $\vec{\gamma}_A$. 

While mutual information is conducive to establishing the viability of some quantum protocols, another metric, concurrence, carries important information in the context of Procrustean filtering procedures. 
Specifically, the average entanglement, that is the product of concurrence and the transmission rate of detected qubits after filtering, remains constant for fixed magnitude of $\gamma_{A}$ and $\gamma_{B}$ \cite{jones2018tuning}.
In other words, there is a tradeoff between the entanglement quality and the transmission rate, which is controlled by the relative orientation of the two filters. 
Mutual information is a monotonic function of concurrence for the cases shown in Figs. 2 (a) and (b), and we observed a corresponding reduction of the transmission rates when the mutual information was restored to higher values, which is in agreement with our previously published results \cite{jones2018tuning}.

Figures \ref{fig:data}(c) and (d) show experimental density matrices taken at various stages of the experiment for the bit-flip and phase-flip cases, respectively. 
In each set, the left-most plots show the initial states of Eqs. \ref{eq:bitflip} and \ref{eq:phaseflip}, which evolve to those shown in the middle plots upon application of the channel A filter $\vec\gamma_{A}$. 
Finally, the right-most plots depict the states after the maximum amount of mutual information was recovered by application of the channel B filter $\vec\gamma_{B}$. 
The same color code is maintained throughout Fig. \ref{fig:data}.

%%%%%%%%%%%%%%%%%%%%%%%%%%%%%%%%%%%%%%%%%%%%%%%%%%%%%%%%%%%%%%%%%%%%%%%%%%%%%%%%%%%%%
\section{Discussion}

%The density matrices in Figs. \ref{fig:data}(c) and (d) provide valuable insight into why the Procrustean entanglement recovery method is ineffective when uncorrelated bit-flip noise, $N_{\text{bf}}$, is mixed into the signal. 
To gain insight into why the Procrustean entanglement recovery method is ineffective when uncorrelated bit-flip noise, $N_{\text{bf}}$, is mixed into the signal, we turn to the density matrices in Figs. 2(c) and (d).
It is illustrative to examine the diagonal matrix elements and the effect that each of the filters, $\vec\gamma_{A}$ and $\vec\gamma_{B}$, has on them. 
Consider partial filters of equal magnitude which are set so that $\vec\gamma_{A}$ preferentially filters out the $|V\rangle$ mode of qubit A and $\vec\gamma_{B}$ preferentially filters out the $|H\rangle$ mode of qubit B. 
These filters counteract each other with respect to the probabilities of detecting co-polarized photons, $|VV\rangle\langle VV|$ and $|HH\rangle\langle HH|$. 
That is, while $\vec\gamma_{A}$ reduces the frequency of simultaneous $|V\rangle$ detections, $\vec\gamma_{B}$ reduces that of simultaneous $|H\rangle$ detections accordingly, so the probabilities of the two events remain equal to each other.  
However, both filters act in concert with respect to the probabilities of detecting cross-polarized photons.
Indeed, both filters favor the $|HV\rangle\langle HV|$ element and reduce the $|VH\rangle\langle VH|$ element.

Uncorrelated bit-flip noise, $N_{\text{bf}}$, has all four co- and cross-polarized diagonal matrix elements.
Therefore, if filters A and B are arranged to restore a $|\phi^{+}\rangle$ Bell state, they can minimize some $N_{\text{bf}}$ elements while adversely increasing others, $|HV\rangle\langle HV|$ in particular. 
Indeed, notice that in the matrix progression of Fig. \ref{fig:data}(c), the matrix element $|HV\rangle\langle HV|$ grows with the application of each filter $\vec\gamma_{A}$ and $\vec\gamma_{B}$. 
This element signifies the presence of cross-polarized detection events, which result in errors for the detection of a $|\phi^{+}\rangle$ Bell state. 
By applying filters A and B, the Bell state component of the overall state (Eq. \ref{eq:bitflip}) can be restored, but the noise component $N_{\text{bf}}$ cannot.
The trade-off between these two components explains why only limited entanglement restoration is possible for bit-flip errors and why $\gamma_{B}^{\text{opt}}<\gamma_{A}$.
On the other hand, correlated noise, $N_{\text{pf}}$, only has pairs of co-polarized photons, so the filters affect the noise in exactly the same way they affect a co-polarized $|\phi^{+}\rangle$ Bell state.
Indeed, the initial state, shown in white on the left-most panel of Fig. \ref{fig:data}(d), can be nearly completely recovered by application of an appropriate filter in channel B as demonstrated by the experimental density matrix shown on the right-most plot of Fig. \ref{fig:data}(d) in cyan. 

%%%%%%%%%%%%%%%%%%%%%%%%%%%%%%%%%%%%%%%%%%%%%%%%%%%%%%%%%%%%%%%%%%%%%%%%%%%%%%%%%%%%%
\section{Conclusion}

To summarize, we transmitted one qubit of an entangled photon pair through a general polarization quantum channel which decoheres and partially filters individual qubit modes. 
In this scenario, the transmitted qubit is impaired by the channel itself, whereas the other qubit remains accessible to a network operator. 
We then studied how the application of a local filter to another qubit maximizes the mutual quantum information between the two. 
We showed that the relative orientation of the decoherence and filtering elements comprising the channel strongly affects the amount of recoverable quantum information. 
We investigated two example cases and established that while only a portion of the lost mutual information can be recovered for bit-flip noise, complete restoration of the mutual information is possible for correlated phase-flip noise. 
We supported the experimental data with corresponding theory and further proposed an elegant intuitive explanation of the observed effects. 
Our results offer tantalizing hope for error-free quantum information transmission in the future.

%%%%%%%%%%%%%%%%%%%%%%%%%%%%%%%%%%%%%%%%%%%%%%%%%%%%%%%%%%%%%%%%%%%%%%%%%%%%%%%%%%%%%
\section{Acknowledgments}
G. Riccardi and C. Antonelli acknowledge financial support from the Italian Government through project INCIPICT and from the U.S. ARO grant W911NF1820155.

%%%%%%%%%%%%%%%%%%%%%%%%%%%%%%%%%%%%%%%%%%%%%%%%%%%%%%%%%%%%%%%%%%%%%%%%%%%%%%%%%%%%%
\beginappendixA
\section{Appendix A. Optimal filter magnitude $\gamma_{B}^{\text{opt}}$}
\label{sec:theoryII}

Here, we expand on our recent theoretical results \cite{kirby2019effect} to derive the magnitude of the filter $\vec\gamma_B$ that produces maximum entanglement. Both states $\rho_{\text{bf}}$ and $\rho_{\text{pf}}$ are Bell diagonal states \cite{horodecki1996information} and can therefore be expressed, up to local unitary rotations, as: 
\begin{equation}
\rho_{\text{BD}}=\frac{1}{4}\left(I_{2}\otimes I_{2}  + \sum_{j=1}^{3}t_{j}\sigma_{j}\otimes\sigma_{j}\right).
\end{equation}
where the $\sigma_{j}$ are the Pauli matrices and $I_{2}$ is the two-dimensional identity matrix, both in Jones space.
The $t_{j}$ are the elements of the diagonal correlation matrix $\textbf{T}=\text{diag}(t_{1},t_{2},t_{3})$ in Stokes space. 
For all rank-2 Bell diagonal states, the $t_j$ coefficients include a single entry $\vert t_{j}\vert=1$ and two $\vert t_{j}\vert=C$, where $C$ is the concurrence of the state $\rho_{\text{BD}}$.
%The application of an arbitrarily oriented local filter of the form $P=e^{-\gamma/2}e^{\frac{1}{2}\vec{\gamma}\cdot\vec{\sigma}}$ to each qubit of $\rho_{BD}$ results in a state $\rho$ with a concurrence of:
The application of an arbitrarily oriented local filter of the form $P=e^{\frac{1}{2}\vec{\gamma}\cdot\vec{\sigma}}$ to each qubit of $\rho_{BD}$ results in a state $\rho$ with a concurrence of:
\begin{equation}
C(\rho)=\frac{C(\rho_{\text{BD}})}{\cosh(\gamma_{A})\cosh(\gamma_{B})+\text{\textbf{T}}\hat{\gamma}_{A}\cdot\hat{\gamma}_{B}\sinh(\gamma_{A})\sinh(\gamma_{B})},
\label{eq:AppendixC}
\end{equation}
%where $\hat{\gamma}_{A}$ and $\hat{\gamma}_{B}$ are the unit Stokes vectors of the local filter elements acting on the first and second qubits respectively, $\text{\textbf{T}}\hat{\gamma}_{A}\cdot\hat{\gamma}_{B}$ is a geometrical dot product, and the operator $\textbf{T}$ is represented by $\textbf{T}=\text{diag}(t_{1},t_{2},t_{3})$.
where $\hat{\gamma}_{A}$ and $\hat{\gamma}_{B}$ are the unit Stokes vectors of the local filter elements acting on the first and second qubits respectively, and $\text{\textbf{T}}\hat{\gamma}_{A}\cdot\hat{\gamma}_{B}$ is a geometrical dot product.
This expression is maximized when the dot product $\text{\textbf{T}}\hat{\gamma}_{A}\cdot\hat{\gamma}_{B}$ is minimized and  $\gamma_{B}^{\text{opt}}=\tanh^{-1}\left(\left\vert\text{Min}\left(\text{\textbf{T}}\hat{\gamma}_{A}\cdot\hat{\gamma}_{B}\right)\right\vert \tanh(\gamma_{A})\right).$

%\begin{equation}
%    \gamma_{B}^{\text{opt}}=\tanh^{-1}\left(\left\vert\text{Min}\left(\text{\textbf{T}}\hat{\gamma}_{A}\cdot\hat{\gamma}_{B}\right)\right\vert \tanh(\gamma_{A})\right).
%\end{equation}
In our experiment, $\hat{\gamma}_{A}$ represents the inherent filtering of channel A, which is inaccesible to the operator. In other words, the value of the dot product $\text{\textbf{T}}\hat{\gamma}_{A}\cdot\hat{\gamma}_{B}$ is restricted such that only rotations of $\hat{\gamma}_{B}$ are permitted. Therefore,  $\left\vert\text{Min}\left(\text{\textbf{T}}\hat{\gamma}_{A}\cdot\hat{\gamma}_{B}\right)\right\vert=\vert\vert \textbf{T}\hat{\gamma}_{A}\vert\vert$.
% In other words, when the Stokes vector of the compensating PDL element is either aligned or anti-aligned with $\textbf{T}\hat{\gamma}_{A}$.
Our two noise scenarios differ by the orientation of the vector $\hat{\gamma}_{A}$ with respect to the basis of the matrix $\textbf{T}$. 
For the phase-flip case, $\hat{\gamma}_{A}$ is aligned with the Stokes axis $\hat{s}_{j}$ for which the element $\vert t_{j}\vert=1$; therefore, $\vert\vert \textbf{T}\hat{\gamma}_{A}\vert\vert=1$, and the optimal magnitude of $\hat{\gamma}_{B}$ is given by $\gamma_{B}^{\text{opt}}=\gamma_{A}$.
On the other hand, for the bit-flip case, $\hat{\gamma}_{A}$ is aligned along a smaller component of $\textbf{T}$ of magnitude $\vert t_j\vert=C$, resulting in an optimal compensating filter magnitude of: 
\begin{equation}
    \gamma_{B}^{\text{opt}}=\tanh^{-1}\left(  C\tanh(\gamma_{A})\right).
\end{equation}

%%%%%%%%%%%%%%%%%%%%%%%%%%%%%%%%%%%%%%%%%%%%%%%%%%%%%%%%%%%%%%%%%%%%%%%%%%%%%%%%%%%%%
\beginappendixB
\section{Appendix B. Mutual information as a function of filter magnitude}
\label{sec:theoryI}

All theoretical values plotted with the lines in Fig. 2 of the main text are the result of numerical extraction of the mutual quantum information from theoretical density matrices.
%Specifically, PDL operators \cite{damask2004polarization} of the form $P=e^{-\gamma/2}e^{\frac{1}{2} \vec{\gamma}\cdot\vec{\sigma}}$ are applied to the initial density matrix $\rho_{\text{in}}$ \cite{kent1999optimal,linden1998purifying,verstraete2001local}:
Specifically, PDL operators \cite{damask2004polarization} of the form $P=e^{\frac{1}{2} \vec{\gamma}\cdot\vec{\sigma}}$ are applied to the initial density matrix $\rho_{\text{in}}$ \cite{kent1999optimal,linden1998purifying,verstraete2001local}:
\begin{equation}
    \rho_{f} = \frac{\left(P_{A}\otimes P_{B}\right) \rho_{\text{in}} \left(P_{A}\otimes P_{B}\right)^{\dagger}}{\text{Tr}\left[\left(P_{A}\otimes P_{B}\right) \rho_{\text{in}} \left(P_{A}\otimes P_{B}\right)^{\dagger}\right]}.
\end{equation}
The initial density matrix $\rho_{\text{in}}$ is set to $\rho_{\text{bf}}$ or $\rho_{\text{pf}}$ with $p=0.33$ for the bit-flip and phase-flip cases, respectively.
Next, the magnitude and alignment of $P_{A}$ and $P_{B}$ are set in accordance with the corresponding experiment.
Finally, the mutual information of the final density matrix $\rho_{f}$ is calculated using $S(A:B) = S(A) + S(B) - S(AB)$.  
%Here, $S(AB)$ is the von Neumann entropy of the entire two-qubit state, $S(\rho_{f}) = -\text{Tr}\{\rho_{f}\,\text{ln}(\rho_{f})\}$, and $S(A)$, $S(B)$ are the von Neumann entropies of the reduced density matrices of qubits A and B, respectively. 
Here, $S(AB)$ is the von Neumann entropy of the entire two-qubit state, $S(\rho_{f}) = -\text{Tr}\{\rho_{f}\,\text{log}_2(\rho_{f})\}$, and $S(A)$, $S(B)$ are the von Neumann entropies of the reduced density matrices of qubits A and B, respectively. 
The theoretical lines were also normalized by a factor of $0.9$. 
This was done so that the theoretical mutual information agreed with the average mutual information of the density matrices measured with $\gamma_{A} = 0$ and $\gamma_{B} = 0$.
Note that a similar normalization factor of $0.925$ was necessary when characterizing the concurrence of a similar entangled photon source in our previous work \cite{jones2018tuning, kirby2019effect}.
We speculate that the additional normalization was required in each experiment due to some unaccounted noise such as Raman scattered photons and/or leakage from the pump. 
%For example, the experimentally generated state shown in Fig. 2(e) of the main text may have some small matrix elements in addition to the Bell state term and the single noise term $N_{\text{bf}}$ included in $\rho_{\text{bf}}$. 
Note that this single normalization factor selected for $\gamma_{A}=0$ and $\gamma_{B}=0$ allows for a nearly perfect fit over the entire range of $\gamma_{A}$ and $\gamma_{B}$ in both the bit-flip and phase-flip scenarios.

%%%%%%%%%%%%%%%%%%%%%%%%%%%%%%%%%%%%%%%%%%%%%%%%%%%%%%%%%%%%%%%%%%%%%%%%%%%%%%%%%%%%%
\section{References}

\bibliography{main}
% Remove the bibliography command and uncomment the following lines before submitting
%\begin{bibliography}
%%%%% COPY CONTENTS OF MY BIBLIOGRAPHY IN HERE
%\end{bibliography}

\end{document}